\documentclass[usenatbib]{mn2e}

\pdfoutput=1 

\usepackage{amsmath, amssymb}
\usepackage{graphicx}
\usepackage{fleqn}
\usepackage{inconsolata}
\usepackage[a4paper,bindingoffset=0.0in,left=0.5in,right=0.5in,top=1in,bottom=1in,footskip=.25in]{geometry}

\newcommand{\E}[1]{\times10^{#1}}

\title[The Thermal Stability of Helium Burning on Accreting Neutron Stars]{The Thermal Stability of Helium Burning on Accreting Neutron Stars}
 
\author[M. Zamfir, A. Cumming \& C. Niquette]{Michael Zamfir\thanks{E-mail: mzamfir@physics.mcgill.ca}$^1$, Andrew Cumming$^1$, Caroline Niquette$^1$\\
$^{1}$ Department of Physics, McGill University, 3600 rue University, Montr\'eal, Qu\'ebec H3A 2T8, Canada}

\begin{document}
\label{firstpage}

\date{Accepted -. Received -; in original form -}
\pagerange{\pageref{firstpage}--\pageref{lastpage}} \pubyear{-}
\maketitle

\begin{abstract}
Thermonuclear burning on the surface of accreting neutron stars is observed to stabilize at accretion rates almost an order of magnitude lower than theoretical models predict. One way to resolve this discrepancy is by including a base heating flux that can stabilize the layer. We focus our attention on pure helium accretion, for which we calculate the effect of a base heating flux on the critical accretion rate at which thermonuclear burning stabilizes. We use the MESA stellar evolution code to calculate $\dot m_{\rm crit}$ as a function of the base flux, and derive analytic fitting formulae for $\dot m_{\rm crit}$ and the burning temperature at that critical accretion rate, based on a one-zone model. We also investigate whether the critical accretion rate can be determined by examining steady-state models only, without time-dependent simulations. We examine the argument that the stability boundary coincides with the turning point $dy_{\rm burn}/d\dot m=0$ in the steady-state models, and find that it does not hold outside of the one-zone, zero base flux case. A linear stability analysis of a large suite of steady-state models is also carried out, which yields critical accretion rates a factor of $\sim3$ larger than the MESA result, but with a similar dependence on base flux. Lastly, we discuss the implications of our results for the ultracompact X-ray binary 4U~1820-30.
\end{abstract}

\begin{keywords}
accretion, instabilities, nuclear reactions -- X-rays: binaries, bursts, individual: 4U~1820-30 -- stars: neutron
\end{keywords}

\section{Introduction}

Thermonuclear burning of hydrogen (H) and helium (He) on the surface of an accreting neutron star is expected to undergo a transition from being thermally-unstable to thermally-stable at a critical accretion rate $\dot M_{\rm crit}\approx 10^{18}\ {\rm g\ s^{-1}}$ (close to the Eddington accretion rate) \citep{Hansen1975,Fujimoto1981}. The transition occurs because the temperature-dependence of the He burning reactions becomes less steep at higher burning temperatures, so that at a high enough accretion rate the reactions are no longer temperature-dependent enough to overcome the stabilizing radiative cooling of the layer. 

Observationally, unstable nuclear burning is seen as Type I X-ray bursts, bright flashes in X-rays with a typical duration of 10--100 seconds that recur on timescales of hours to days \citep{Lewin1993}. Consistent with the idea that the burning stabilizes, the rate of Type I X-ray bursts drops dramatically in several sources above a persistent luminosity $L_X\approx 2\times 10^{37}\ {\rm erg\ s^{-1}}$ (\citealt{Cornelisse2003}; see also \citealt{Clark1977}, citealt{vanParadijs1988}), and the burst energetics clearly point to most of the accreted fuel burning in a stable manner \citep{vanParadijs1988,Galloway2008}. Other observed phenomena also point to stable burning at high accretion rates. Stable H/He burning is required in models for superbursts to produce the carbon fuel that is believed to drive those events \citep{Schatz2003,Woosley2004,Stevens2014}, and is manifested in the energetics of Type I X-ray bursts observed from superburst sources \citep{intZand2003}. The mHz QPOs observed in some sources \citep{Revnivtsev2001,Altamirano2008,Linares2012} have been identified with an oscillatory mode of nuclear burning that emerges when the burning is marginally-stable, i.e.~transitioning between stable and unstable \citep{Paczynski1983a, NarayanHeyl2003, Heger2007QPOs, Keek2014}.

Despite this qualitative agreement, a long-standing puzzle has been that the observed accretion rate at which the onset of stable burning occurs is $\dot M\sim 10^{17}\ {\rm g\ s^{-1}}$, an order of magnitude lower than theory predicts given standard assumptions for the thermal state of the crust \citep{Brown2000, Bildsten2000, Keek2014}. Several mechanisms have been suggested to account for this discrepancy, including a change in burning mode to slowly propagating fires around the neutron star surface \citep{Bildsten1995}, partial covering of the accreted fuel \citep{Bildsten1998}, mixing of fuel driven by rotational instabilities \citep{Fujimoto1987,Piro2007,Keek2009}, and strong heating of the layer associated with spin-down and spreading of the fuel following disk accretion \citep{Inogamov1999, Inogamov2010}, or other sources of heating \citep{Bildsten1995, NarayanHeyl2002, NarayanHeyl2003, Keek2009}.

The proposal that the unstable burning is quenched by heating is intriguing because evidence has accumulated that the outer crust and ocean of accreting neutron stars are strongly heated by an unknown shallow heat source. One piece of evidence is from superbursts, whose observed ignition properties require temperatures of $\approx 6\times 10^8\ {\rm K}$ be achieved at column depths of $\approx 10^{12}\ {\rm g\ cm^{-2}}$ in the neutron star ocean, requiring an additional source of heat be added to models \citep{Brown2004, Cumming2006}. This problem has been exasperated recently with observations of superbursts in transient systems \citep{Keek2008, Altamirano2012}. The second piece of evidence is from modelling of the thermal relaxation of transiently-accreting neutron stars in quiescence. \cite{Brown2009} found that the temperatures observed in KS~1731-260 and MXB~1659-29 approximately one month into quiescence required an inwards heat flux into the neutron star crust and a corresponding strong shallow heat source. \citet{Degenaar2013} reached a similar conclusion based on rapid cooling of XTE~J1709-267 after a short 10 week outburst. \cite{Schatz2014} showed that a strong neutrino cooling source may operate in the outer crust, emphasizing the need for additional heating at shallow depths. Finally, modelling of X-ray burst recurrence times in a number of sources has suggested that outwards fluxes of $\sim 0.3\ {\rm MeV}$ per nucleon\footnote{Throughout the paper we will measure the heat flux in units of the equivalent energy per accreted nucleon $Q_b$ in MeV per nucleon, so that the flux is $F=Q_b \dot m$, where $\dot m$ is the local accretion rate $\dot m = \dot{M} / 4 \pi R^2$.} or more heat the accumulating H/He layer \citep{Cumming2003,Galloway2006}.

Determining the dependence of $\dot M_{\rm crit}$ on the base flux is critical to assess whether shallow heating could also be the reason for stabilization of Type I X-ray bursts at observed accretion rates $\dot M\gtrsim  10^{17}\ {\rm g\ s^{-1}}$. Most calculations of the critical accretion rate $\dot M_{\rm crit}$ in the literature are for a fixed base flux, typically $Q_b\approx 0.1\ {\rm MeV}$ per nucleon (taken from models of the global thermal state of the neutron star, e.g.~\citealt{Brown2000}) for which $\dot M_{\rm crit}\approx 10^{18}\ {\rm g\ s^{-1}}$ (e.g. \citealt{Heger2007QPOs,Keek2014}). \cite{Bildsten1995} calculated the effect of a flux from deep carbon burning on the stability of the helium shell using a one-zone approach and \citet{Fushiki1987} also included the base temperature as a parameter in their one-zone study. Using a linear stability analysis of models with fixed temperatures set below the accreted layer, \citet{NarayanHeyl2002, NarayanHeyl2003} found transitions to stable burning for a solar mixture of hydrogen and helium at accretion rates of around $\sim0.1-0.3\,\dot{M_{\rm Edd}}$. \cite{Keek2009} calculated the stability boundary for pure helium accretion using detailed multizone models for several different base fluxes, showing that an increased heating rate decreases $\dot M_{\rm crit}$. They found that a base luminosity of $L_{\rm crust}\approx 10^{35}\ {\rm erg\ s^{-1}}$ (approximately $1\ {\rm MeV}$ per nucleon at 0.1 Eddington) lowered the critical accretion rate to $\approx 10^{17}\ {\rm g\ s^{-1}}$. Analogous simulations varying base flux for H/He accretion have not been carried out. When the accreted material contains a significant amount of hydrogen, the burning proceeds via the rp-process involving hundreds of nuclei \citep{WallaceWoosley1981} and so calculations are much more numerically-intensive and so far have been carried out only for specific choices of base flux \citep{Schatz1998,Woosley2004,Keek2014}.

In this paper, we take some further steps towards calculating and understanding the variation of $\dot M_{\rm crit}$ with base flux. For simplicity, we consider only pure helium accretion, but with the goal of developing techniques that can be readily applied to the mixed H/He accretion case later. We first use the stellar evolution code MESA \citep{Paxton2011,Paxton2013} to confirm the results of \cite{Keek2009} for pure helium accretion. We then extend the one-zone model of \cite{Bildsten1998} to include a base flux, which we use to understand the shape of the relation between $\dot M_{\rm crit}$ and $Q_b$, and to derive useful fitting formulae. In the second part of the paper, we investigate two different methods that have been proposed to determine the stability of the nuclear burning based purely on a steady-state model at a given accretion rate, rather than running time-dependent simulations. This is potentially very powerful because steady-state models can be calculated quickly even when rp-process burning is included (e.g.~see the large grid of steady-state models recently calculated by \citealt{Stevens2014}).

An outline of the paper is as follows. The time-dependent simulations of helium accretion and one-zone analysis are presented in \S 2. In \S 3, we discuss the relation between the burning depth in steady-state models and the thermal stability of the model. In \S 4, we develop a linear stability analysis of steady-state models and compare to the time-dependent results from MESA. We conclude in \S 5, where we also discuss the application of our results to the ultracompact X-ray binary 4U~1820-30.


\section{The effect of base heating on the stability boundary}

We start in this section by calculating the critical accretion rate $\dot M_{\rm crit}$ for pure helium accretion as a function of the base flux $Q_b$. The results of our time-dependent simulations are presented in \S 2.1, and a one-zone model is developed in \S 2.2 to help to understand the results.

\begin{figure}
\includegraphics[width=0.5\textwidth]{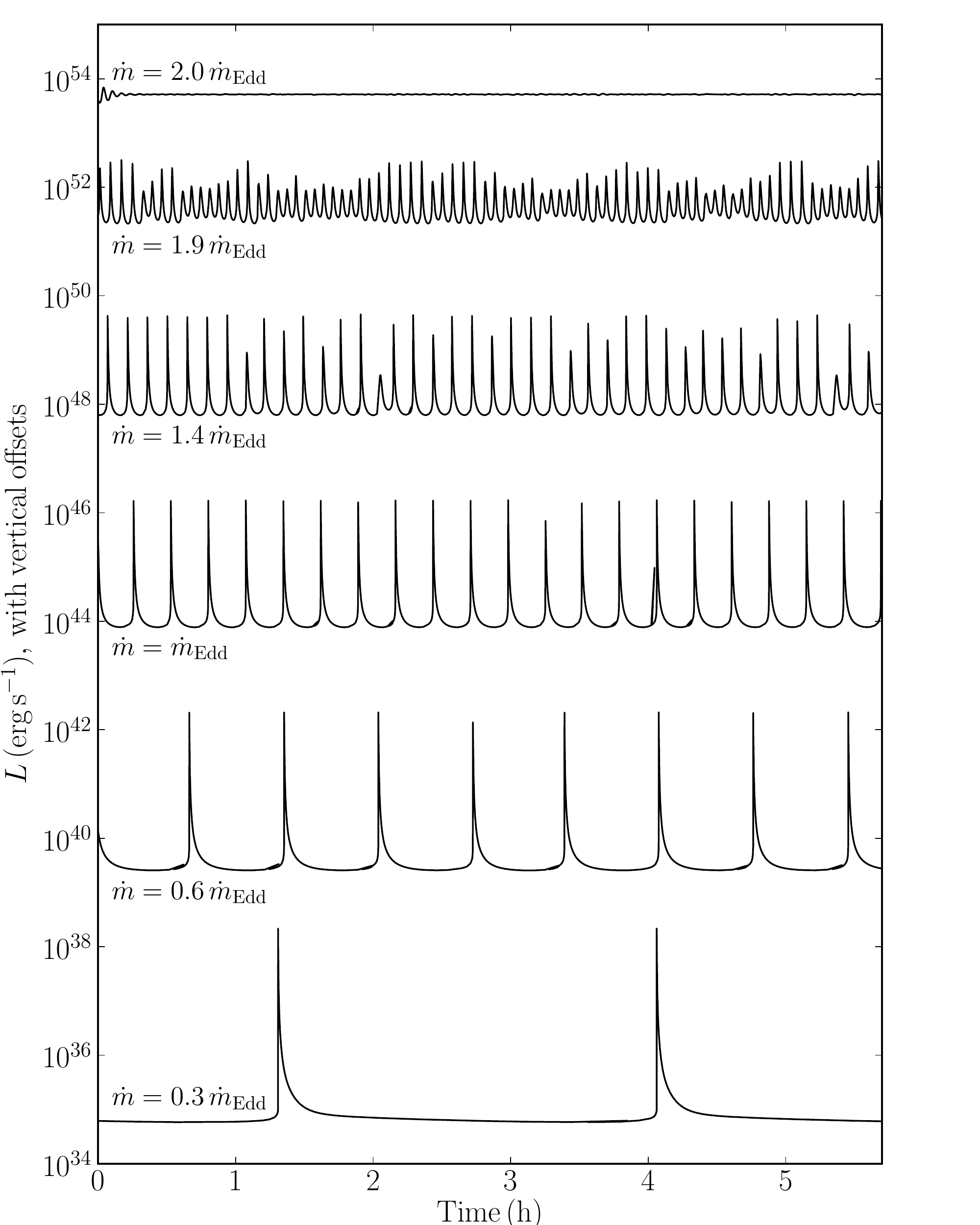}
\caption{Lightcurve profiles generated by MESA showing stable burning and various bursting behaviours at different accretion rates. All models are computed with a base flux of $Q_b=0.1$\,MeV\,nuc$^{-1}$. In going from $\dot{m}=1.9\dot{m}_{\rm Edd}$ to $\dot{m}=2.0\dot{m}_{\rm Edd}$, the amplitude abruptly changes from $\Delta L/L\approx10$ to roughly zero, illustrating the fact that modest variations to the instability criterion we used ($\Delta L/L>2$) do not strongly affect the location of the stability boundary.
\label{fig:mesa_regimes}}
\end{figure}

\subsection{Time dependent calculations with MESA}\label{sec:mesa}

One of the exciting developments in stellar astrophysics in recent years has been the release of the open source stellar evolution code MESA (Modules for Experiments in Stellar Astrophysics) \citep{Paxton2011,Paxton2013}. MESA solves the equations of stellar evolution in a fully-coupled way, and includes the relevant microphysics for the outer layers of a neutron star relevant for Type I X-ray bursts. Indeed, a sequence of helium flashes on an accreting neutron star was modelled in \cite{Paxton2011}, and accretion onto a neutron star is a standard test case in the MESA distribution. We apply MESA here to determine the stability boundary for pure helium accretion. We view this as a straightforward first step to developing MESA as a general tool to study X-ray bursts on accreting neutron stars. Here, we will present only our results on the stability boundary, leaving a detailed analysis of burst sequences and the evolution of the burning layers during a burst for a future paper.

We used the MESA release 6596 for our simulations. To enable a meaningful comparison with one-zone models and linear stability analysis (\S 4), we used a simplified nuclear network that takes into account only the triple alpha reaction $3\alpha\rightarrow ^{12}$C, so that only two species, He and carbon, were present. For determining the stability boundary, this is in fact a good approximation: we also tried using the approx21 network that includes a sequence of helium burning reactions to heavier elements, and found that the critical accretion rate changed by $\lesssim 10$\% with the change of network. The reason for this is that the burning temperature at the stability boundary, $T\approx 3$--$4\times 10^8\ {\rm K}$, is small enough that the burning does not proceed 
significantly past carbon (e.g.~\citealt{Brown1998}).

From this point onwards, we will use local values for the accretion rate. We adopt a standard value for the local Eddington accretion rate, $\dot m_{\rm Edd}=8.8\times 10^4\ {\rm g\ cm^{-2}\ s^{-1}}$ (the equivalent global accretion rate is $\dot M_{\rm Edd}=1.11\times 10^{18}\ {\rm g\ s^{-1}} = 1.74\times 10^{-8}\ M_\odot \ {\rm yr^{-1}}$). This corresponds to the Eddington rate for solar composition; we use it here as a standard value even though our simulations are for pure helium accretion. We assume a $1.4\ M_\odot$, $10\ {\rm km}$ neutron star, which has a surface gravity of $g=1.9\E{14}$\,cm\,s$^{-2}$. This is the Newtonian value for the surface gravity, and does not include the general relativistic correction; however the dependence of the critical accretion rate on gravity is weak (see \S\ref{sec:results}).

To find the critical accretion rate, we followed these steps. For each choice of $Q_b$ and $\dot m$, we first accrete a column $10^{10}\ {\rm g\ cm^{-2}}$ of carbon, allowing the model to thermally adjust to the base luminosity. We then accrete an additional column of $10^{10}\ {\rm g\ cm^{-2}}$ of pure helium. Since the burning depth is $\sim 10^8\ {\rm g\ cm^{-2}}$, this means that we accrete a column of roughly one hundred burning depths which allows the initial transient behaviour to die away at the beginning of the run. We then assess whether the burning has stabilized by looking at the range of luminosities in the last 10\% of the lightcurve. If the luminosity variation is smaller than a factor of $\Delta L/L=2$ then we classify the burning as stable. We have checked that our derived stability boundary does not significantly change if we use another value for $\Delta L/L$. To find convergence in the stability boundary, we used a value of 0.1 for the parameter \texttt{mesh\textunderscore delta\textunderscore coeff} in the MESA code. We found that using a value ten times larger for this parameter yielded critical accretion rates that differed by $\sim10\%$. For each $Q_b$, we start at a large accretion rate and run successive models with accretion rate reduced in steps of $\Delta \log_{10}\dot m/\dot{m}_{\rm Edd}=0.025$, until the burning becomes unstable, which means that we have located the stability boundary. Figure \ref{fig:mesa_regimes} shows that, for a base flux $Q_{b}=0.1$\,MeV per nucleon, the initially stable behaviour transforms to a sequence of bursts below $2\,\dot m_{\rm Edd}$, with the burst recurrence time and amplitude growing as the accretion rate is lowered further.

The stability boundary as a function of base flux is shown in Figure 2. We see a smooth decrease in $\dot m_{\rm crit}$ with $Q_b$, reaching $0.1\ \dot m_{\rm Edd}$ at $Q_b\approx 0.7$ MeV. The results of \cite{Keek2009} are shown as a comparison (note that \citealt{Keek2009} present these results as luminosity against $\dot m$, see their Figure 11, here we have divided the luminosity by $\dot m$ to convert the luminosity to MeV per nucleon units). The agreement is good, with typical deviations of tens of percent, although the point at $Q_b\approx 0.8$\,MeV from \cite{Keek2009} is a factor of 2 higher than the MESA result.

\subsection{One zone model}\label{sec:onezone}

To understand the shape of the $\dot m_{\rm crit}(Q_b)$ relation, it is helpful to consider a one-zone model with a base flux included. In a one-zone treatment of the burning layer, we follow the layer temperature and column depth according to
\begin{eqnarray}\label{eq:dTdt}
c_P \frac{d T}{dt}\ &=&\ \epsilon_{3\alpha} - \epsilon_\mathrm{cool} + \frac{Q_b \dot{m}}{y},\\\label{eq:dydt}
\frac{dy}{dt}\ &=&\ \dot{m} -\frac{\epsilon_{3\alpha}}{E_{3\alpha}}y
\end{eqnarray}
\citep{Paczynski1983a,Bildsten1998,Heger2007QPOs}, where $\epsilon_{3\alpha}$ is the heating rate and $E_{3\alpha}$ is the energy per unit mass released from $3\alpha$ reactions, and the one-zone cooling rate is $\epsilon_{\rm cool}\approx acT^4/3\kappa y^2$. Heating from beneath the layer is represented by the third term on the right side of equation (\ref{eq:dTdt}) \citep{Heger2007QPOs}. 

To derive the stability boundary, we consider steady-state solutions of equations (\ref{eq:dTdt}) and (\ref{eq:dydt}) and perturb them, taking the perturbations to be at constant pressure and column depth (since column depth $y=P/g$ in a thin layer). We follow \cite{Bildsten1998} and assume an ideal gas equation of state, so that $\delta\rho/\rho = -\delta T/T$ at constant pressure. This gives
\begin{flalign}
c_P{\frac{d\delta T}{dt}} &= \delta \epsilon_{3\alpha}-\delta \epsilon_\mathrm{cool}\notag\\
&={\delta T\over T} \left[ \epsilon_{3\alpha}\left(\nu-\eta\right) - \epsilon_{\rm cool} \left(4-\kappa_T\right) \right]
\end{flalign}
where we have expressed the heating rate as $\epsilon_{3\alpha}\propto \rho^\eta T^\nu$, and $\kappa_T=\left.\partial \ln\kappa/\partial \ln T\right|_P$. For triple alpha burning, $\nu\approx (44/T_8)-3$ and $\eta=2$, where $T_8=T/10^8$ K \citep{HansenKawaler1994}.

\begin{figure}
\includegraphics[width=0.5\textwidth]{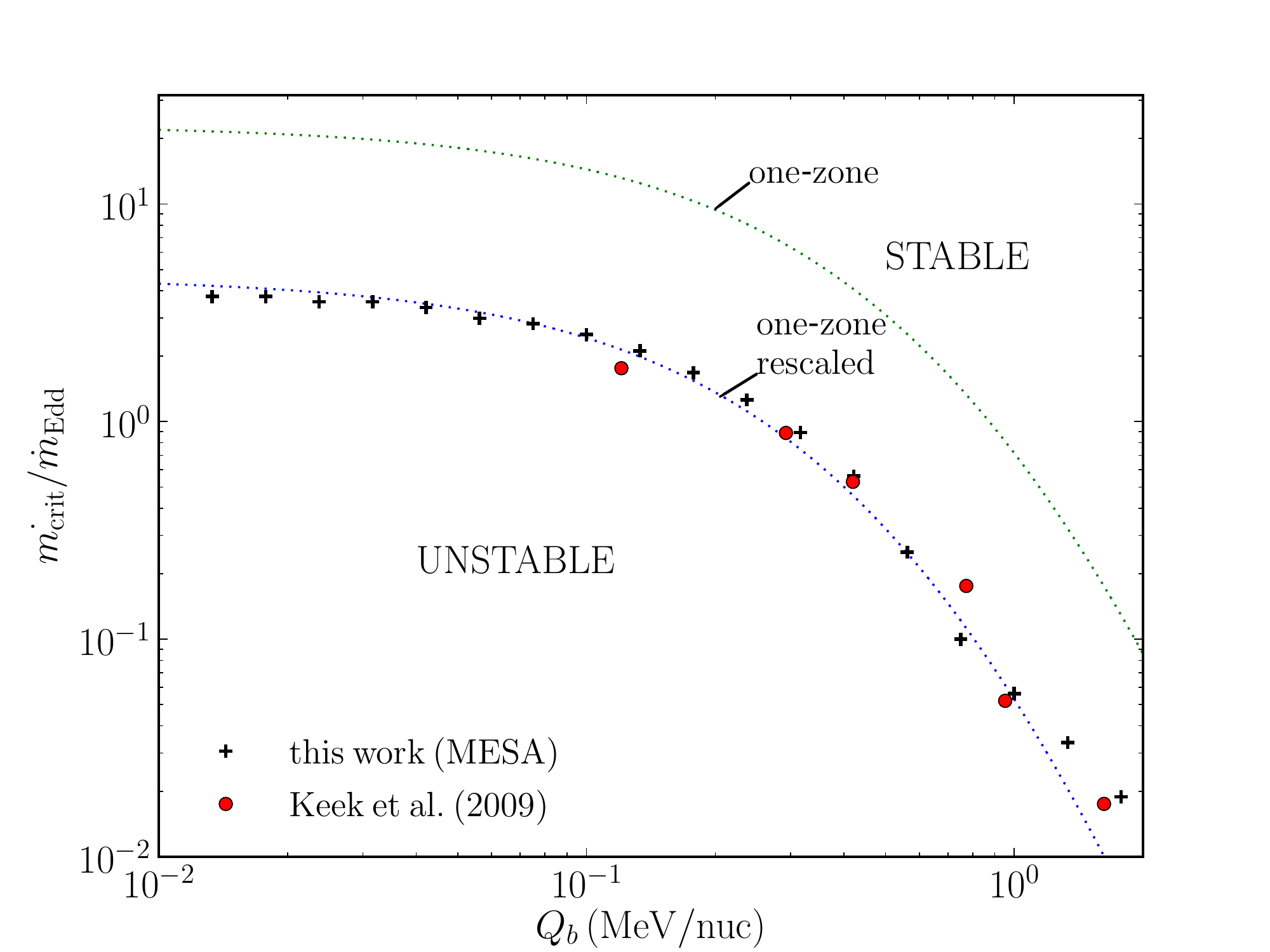}
\caption{The stability boundary we find using MESA (black crosses) agrees well with the boundary found by \citet{Keek2009} (red circles). The analytically-derived one-zone estimate to the stability boundary (equation \ref{eq:onezone_mdotcrit}) is shown as a green dotted curve. The one-zone curve shares a similar shape to, but overestimates the values of the MESA data. By making small adjustments to equation (\ref{eq:onezone_mdotcrit}) (see text), we obtain an analytic fit to the MESA results (equation \ref{eq:mdotcrit}), shown as the blue dotted curve. 
\label{fig:onezone_all}}
\end{figure}

When the base heat flux is much smaller than the energy generated inside the layer, the steady-state obeys $\epsilon_{3\alpha}\approx \epsilon_{\rm cool}$, and the condition for instability ($d\delta T/dt>0$) is
\begin{equation}\label{eq:instability_criterion_Qb0}
\nu-\eta-4+\kappa_T = {\frac{44}{T_8}}-9+\kappa_T >0
\end{equation}
(e.g.~\citealt{Bildsten1995,Bildsten1998,Yoon2004}).
When the base flux is significant, $\epsilon_{3\alpha}$ is no longer equal to $\epsilon_{\rm cool}$, and in fact is smaller since the base flux $Q_b$ now contributes to the heating of the layer. The instability condition is 
\begin{equation}\label{eq:instability_criterion}
\nu-\eta-{\epsilon_{\rm cool}\over \epsilon_{3\alpha}}\left(4-\kappa_T\right)>0.
\end{equation}
In steady-state, equation (\ref{eq:dTdt}) gives $\epsilon_{\rm cool}=\epsilon_{3\alpha}+\dot m Q_b/y$, and the burning depth is given by equation (\ref{eq:dydt}) as $y/\dot m = E_{3\alpha}/\epsilon_{3\alpha}$. Therefore
\begin{equation}\label{eq:ecoolovere3a}
{\epsilon_{\rm cool}\over \epsilon_{3\alpha}}=1+{Q_b\over E_{3\alpha}}=1+{Q_b\over 0.61\ {\rm MeV}}.
\end{equation}
We see that when $Q_b$ is significant, the cooling term in the instability criterion is enhanced. This implies that to trigger a burst in the presence of a base flux, the temperature in the layer must be lower than without the base flux, so that $\nu$ is larger and able to overcome the cooling term. The effect of $Q_b$ is therefore to lower $\dot m_{\rm crit}$ compared to its $Q_b=0$ value, as seen in the MESA results in Figure 2.

\begin{figure}
\includegraphics[width=0.5\textwidth]{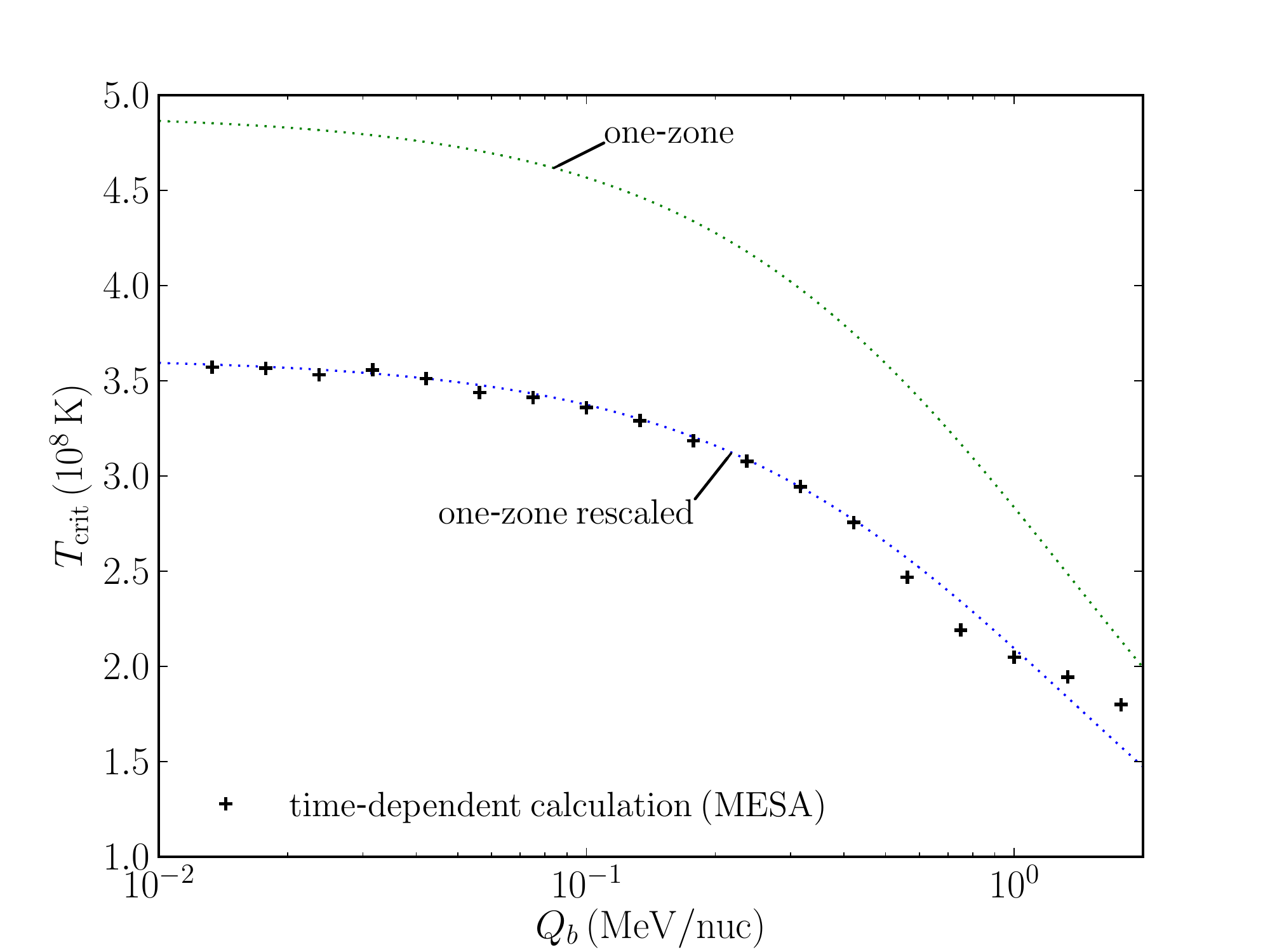}
\caption{The temperature at the triple-alpha burning depth along the MESA stability boundary is shown in black crosses. Sharing a similar shape but with higher values, the one-zone analytical estimate to the critical temperature (equation \ref{eq:analTb}) is shown as a dotted green line. By changing the prefactor in equation (\ref{eq:analTb}) from 4.9 to 3.5, we find a good fit to the MESA data, allowing us to establish a simple analytical expression for the burning temperature at the critical accretion rate.
\label{fig:Tb}}
\end{figure}

Setting the equality in equation (\ref{eq:instability_criterion}) gives an expression for the critical temperature below which helium burning becomes unstable in the one-zone model,
\begin{equation}\label{eq:analTb}
T_{\rm{crit,}\,8} = 4.9\ \left[{5\over 9}+{4-\kappa_T\over 9}\left(1+{Q_b\over 0.61\ {\rm MeV}}\right)\right]^{-1}.
\end{equation}
Setting $Q_b$ and $\kappa_T$ to zero, we recover the critical temperature for stable helium burning found by \citet{Bildsten1998}, $T_{\rm{crit,}\,8}\,=\,4.9$. Equation (\ref{eq:analTb}) confirms that the inclusion of a base heating flux reduces the critical temperature required for the onset of unstable burning. As \citet{Keek2009} noticed, $T_8=4.9$ is well above the burning temperature at marginal stability in multizone time-dependent calculations (see Figure \ref{fig:Tb}). However since the shapes of the one-zone and MESA curves agree well, we can adjust the prefactor in equation (\ref{eq:analTb}) from 4.9 to 3.6 to obtain a simple analytical expression that describes the burning temperature at the critical accretion rate in MESA (shown as a blue dotted curve in Figure \ref{fig:Tb}).

We can now find $\dot m_{\rm crit}$ at a given $Q_b$ by calculating the accretion rate at which the burning temperature is equal to the value in equation (\ref{eq:analTb}). To do so, we can use the following expression (\citealp{Bildsten1998}, equation 19) which gives the temperature at the helium burning depth in steady-state,
\begin{equation}\label{eq:BildTburn}
T_{\rm burn}=2.8\times 10^8\ {\rm K}\ \left({\dot m\over \dot m_{\rm Edd}}\right)^{1/5} \left(1+{Q_b\over 0.61\ {\rm MeV}}\right)^{3/20},
\end{equation}
where we assume pure helium composition and other appropriate parameters ($\mu=4/3$, $E_{18}=0.58$, $g_{14}=1.9$) and have written the flux heating the layer in terms of $Q_b$. The scalings in this expression indicate that a reduction in the critical temperature required for instability implies a reduction in the accretion rate, for a constant base flux. Equating the temperatures in equations (\ref{eq:analTb}) and (\ref{eq:BildTburn}), we find the critical accretion rate
\begin{equation}\label{eq:onezone_mdotcrit}
\dot m_{\rm crit} = 16\ \dot m_{\rm Edd}\ \left(1+{Q_b\over 0.61\ {\rm MeV}}\right)^{-3/4}\left(1+{Q_b\over 1.37\ {\rm MeV}}\right)^{-5},
\end{equation}
where we again set $\kappa_T=0$. 

We have checked equation (\ref{eq:onezone_mdotcrit}) by running time-dependent one zone models, solving equations (\ref{eq:dTdt}) and (\ref{eq:dydt}) in time. We include electron scattering, free-free, and conductive opacities following \cite{Schatz1999} and \cite{Stevens2014}, the 3$\alpha$ burning rate from \citet{FushikiLamb1987}, and we used fitting formulae for the contributions of degenerate and relativistic electrons to the equation of state from \citet{Paczynski1983b}. We use a similar method to the MESA runs described in \S 2.1 to determine from the lightcurve whether the burning is stable or unstable. We find that the analytic expression in equation (\ref{eq:onezone_mdotcrit}) underestimates the time-dependent one-zone $\dot m_{\rm crit}$ by 30--50\% across the range of $Q_b$. These differences are mostly due to the assumptions of constant opacity $\kappa=0.136\ {\rm cm^2\ g^{-1}}$ and ideal gas equation of state that go into equation (\ref{eq:BildTburn}). We have confirmed this by running time-dependent models that adopt the same assumptions. At low accretion rates and fluxes $Q_b\gtrsim 1$ MeV, another source of error is that the approximation $\exp(-44/T_8)\approx 2.22\times 10^{-6}(T_8/3.38)^{13}$ used by \cite{Bildsten1998} to expand the triple alpha burning rate as a power law begins to break down.

We find that adjusting the prefactor in equation (\ref{eq:onezone_mdotcrit}) to $23\ \dot m_{\rm Edd}$ reproduces the full time-dependent one-zone model to within $10-20$\%, and this is plotted as a green dotted curve in Figure \ref{fig:onezone_all}. 

\subsection{Analytic expression for $\dot m_{\rm crit}(Q_b)$}

Comparing the one-zone result with the MESA calculation in Figure \ref{fig:onezone_all} shows that the overall shape of the curve is reproduced well by the one-zone model, but the magnitude of $\dot m_{\rm crit}$ is overestimated by a factor of approximately 5. This factor is similar to the difference between the $\dot m_{\rm crit}=16\ \dot m_{\rm Edd}$ found by \cite{Bildsten1995} and the $\dot m_{\rm crit}\approx 3\ \dot m_{\rm Edd}$ found by \cite{Keek2009}. The inaccuracy of the one-zone model comes from the instability criterion equation (\ref{eq:instability_criterion}) which overestimates the critical temperature for stable burning (eq.~[\ref{eq:analTb}]). Equation (\ref{eq:BildTburn}) for the burning temperature of the layer, which comes from an analytic integration of the temperature profile in the layer, is quite accurate. For example, at the low flux stability boundary $\dot m=3.5\ \dot m_{\rm Edd}$, equation (\ref{eq:BildTburn}) predicts $T_8=3.6$ which agrees well with the burning temperature (see Fig.~\ref{fig:Tb}).

To obtain an analytic fit to the MESA results, we rescaled equation (\ref{eq:onezone_mdotcrit}) by adjusting the prefactor and making a small adjustment to the numerical constant inside the final term to improve the fit at intermediate values of $Q_b$. The final result, shown in Figure \ref{fig:onezone_all} as a blue dotted curve, is
\begin{align}\label{eq:mdotcrit}
\dot m_{\rm crit} = 4.6\ &\dot m_{\rm Edd}\ \left(1+{Q_b\over 0.61\ {\rm MeV}}\right)^{-3/4}\notag\\
&\times\left(1+{Q_b\over 0.9\ {\rm MeV}}\right)^{-5},
\end{align}
which reproduces the MESA results to within $\lesssim 18$\% for $Q_b\leq 1$ MeV.


\section{The relation between the steady-state burning depth and stability}\label{sec:dydmdot}

In this section we investigate the relation between the burning depth in steady-state models and thermal stability. \cite{Paczynski1983a} pointed out that in one-zone models, the burning depth $y_{\rm burn}$ decreases with $\dot m$ for unstable models ($dy_{\rm burn}/d\dot m<0$), but increases with $\dot m$ in stable models ($dy_{\rm burn}/d\dot m>0$). The stability boundary is therefore at the turning point $dy_{\rm burn}/d\dot m=0$. \cite{NarayanHeyl2003} argued that the same criterion should apply to multizone models also. If this result is generally true, it would be a very powerful way to determine the stability boundary without doing any time-dependent calculations, and large grids of steady-state models already exist as functions of $\dot m$, $Q_b$ and accreted composition (helium fraction) \citep{Stevens2014}. 

We first discuss the one-zone case in \S 3.1, extending the arguments of \cite{Paczynski1983a} to the case with $Q_b>0$. We then consider multizone models in \S 3.2. We show that in both cases $dy_{\rm burn}/d\dot m$ is non-zero at marginal stability, and so $dy_{\rm burn}/d\dot m=0$ can be used to locate the marginally stable point only for one-zone models with $Q_b=0$.

\subsection{The turning point and stability of one-zone models}\label{sec:dydmdot_1zone}

First consider the case studied by \cite{Paczynski1983a}, a sequence of one-zone models with increasing $\dot m$, and $Q_b=0$. From equations (\ref{eq:dTdt}) and (\ref{eq:dydt}), these models must obey 
\begin{eqnarray}\label{eq:ss1}
\epsilon_{3\alpha}&=&\epsilon_{\rm cool},\\
\label{eq:ss2}\dot m &=& {\epsilon_{3\alpha}\over E_{3\alpha}}y,
\end{eqnarray}
in steady-state. At the accretion rate where $dy_{\rm burn}/d\dot m=0$, two neighbouring steady-state models which differ in accretion rate by an amount $\Delta\dot m$ have the same burning depth, so $\Delta y_{\rm burn}=0$. Equation (\ref{eq:ss2}) then gives $\Delta \dot m = (y/E_{3\alpha})\Delta \epsilon_{3\alpha}$. Since the column depth remains unchanged, the difference in accretion rates between the two models is accommodated by a change in the burning rate driven by a temperature difference at constant pressure (column depth), $\Delta \epsilon_{3\alpha} = (\nu-\eta) \epsilon_{3\alpha}\Delta T/T$. The temperature difference between the two models also implies a difference in cooling rates $\Delta\epsilon_{\rm cool}=(4-\kappa_T)\epsilon_{\rm cool}\Delta T/T$, and so setting $\Delta\epsilon_{3\alpha}=\Delta\epsilon_{\rm cool}$ as must be the case for two steady-state models, we arrive at
\begin{equation}
\left(\nu-\eta\right)=\left(4-\kappa_T\right),
\end{equation}
exactly the criterion for marginal stability (see eq.~[\ref{eq:instability_criterion_Qb0}]). Therefore, we have shown that the steady-state model with $dy_{\rm burn}/d\dot m=0$ is marginally stable.

When a base flux is included, equation (\ref{eq:ss1}) becomes
\begin{equation}\label{eq:ss3}
\epsilon_{3\alpha}+{Q_b\dot m\over y} = \epsilon_{\rm cool}.
\end{equation}
Two neighbouring models at $dy_{\rm burn}/d\dot m=0$ are still related by $\Delta \dot m = (y/E_{3\alpha})\Delta \epsilon_{3\alpha}$ because equation (\ref{eq:ss2}) has not changed, but from equation (\ref{eq:ss3}), they must now satisfy
\begin{equation}
\Delta \epsilon_{3\alpha}+{Q_b\over y}\Delta\dot m =\Delta \epsilon_{3\alpha} \left( 1 +{Q_b\over E_{3\alpha}}\right)= \Delta\epsilon_{\rm cool} 
\end{equation}
or
\begin{equation}
\epsilon_{3\alpha}\left( 1 +{Q_b\over E_{3\alpha}}\right)\left(\nu-\eta\right)-\epsilon_{\rm cool}\left(4-\kappa_T\right)=0.
\end{equation}
But the steady-state model obeys $\epsilon_{3\alpha}(1+Q_b/E_{3\alpha})=\epsilon_{\rm cool}$ (eq.~[\ref{eq:ecoolovere3a}]), giving again
\begin{equation}\label{eq:ss4}
\left(\nu-\eta\right)=\left(4-\kappa_T\right)
\end{equation}
at the accretion rate where $dy_{\rm burn}/d\dot m=0$. But for $Q_b>0$ this is no longer the condition for marginal stability (see eq.~\ref{eq:instability_criterion}). Therefore the turning point for $y_{\rm burn}$ no longer specifies the stability boundary when $Q_b>0$.

It is curious that at the turning point for any value of $Q_b$, the criterion for marginal stability at $Q_b=0$ (eq.~[\ref{eq:ss4}]) is satisfied. This means that models where $dy_{\rm burn}/d\dot m=0$ for any value of $Q_b$ will have the same burning temperature $T_8=4.9$ at which equation (\ref{eq:ss4}) is satisfied.

\subsection{The turning point and stability of multizone models}\label{sec:steady}

To locate the turning point $dy_{\rm burn}/d\dot m=0$ in the multizone case, we constructed a set of steady-state models of the helium burning layer as a function of $Q_b$ and $\dot m$. We solve for the temperature $T$, helium mass fraction $Y$, and flux $F$ as a function of column depth $y$ by integrating (see \citealt{Brown1998})
\begin{equation}\label{eq:dTdy}
{dT\over dy} = {3\kappa F\over 4acT^3}
\end{equation}
\begin{equation}\label{eq:dFdy}
{dF\over dy} = -\epsilon_{3\alpha} + {\dot m c_PT\over y}\left(\nabla-\nabla_{\rm ad}\right)
\end{equation}
and
\begin{equation}\label{eq:dYdy}
{dY\over dy} = -{12\epsilon_{3\alpha}\over \dot m Q_{3\alpha}},
\end{equation}
where $Q_{3\alpha}=7.275$ MeV is the energy release from one triple-alpha reaction. We assume that helium burns to carbon only, with the triple-alpha generation rate $\epsilon_{3\alpha}$, opacity $\kappa$ and equation of state calculated in the same way as in \S\ref{sec:onezone}. The boundary conditions are $Y=1$ at the top of the layer, and $F=Q_b\dot m$ at the base. The flux at the top has contributions from $Q_b$, the nuclear burning, and the compressional heating, described by the term involving $\nabla_{\rm ad}-\nabla$ on the right hand side of equation (\ref{eq:dFdy}). Since the compressional heating depends on the temperature profile, it is necessary to iterate the solution until the assumed compressional heating is self-consistent. 

\begin{figure}
\includegraphics[width=0.5\textwidth]{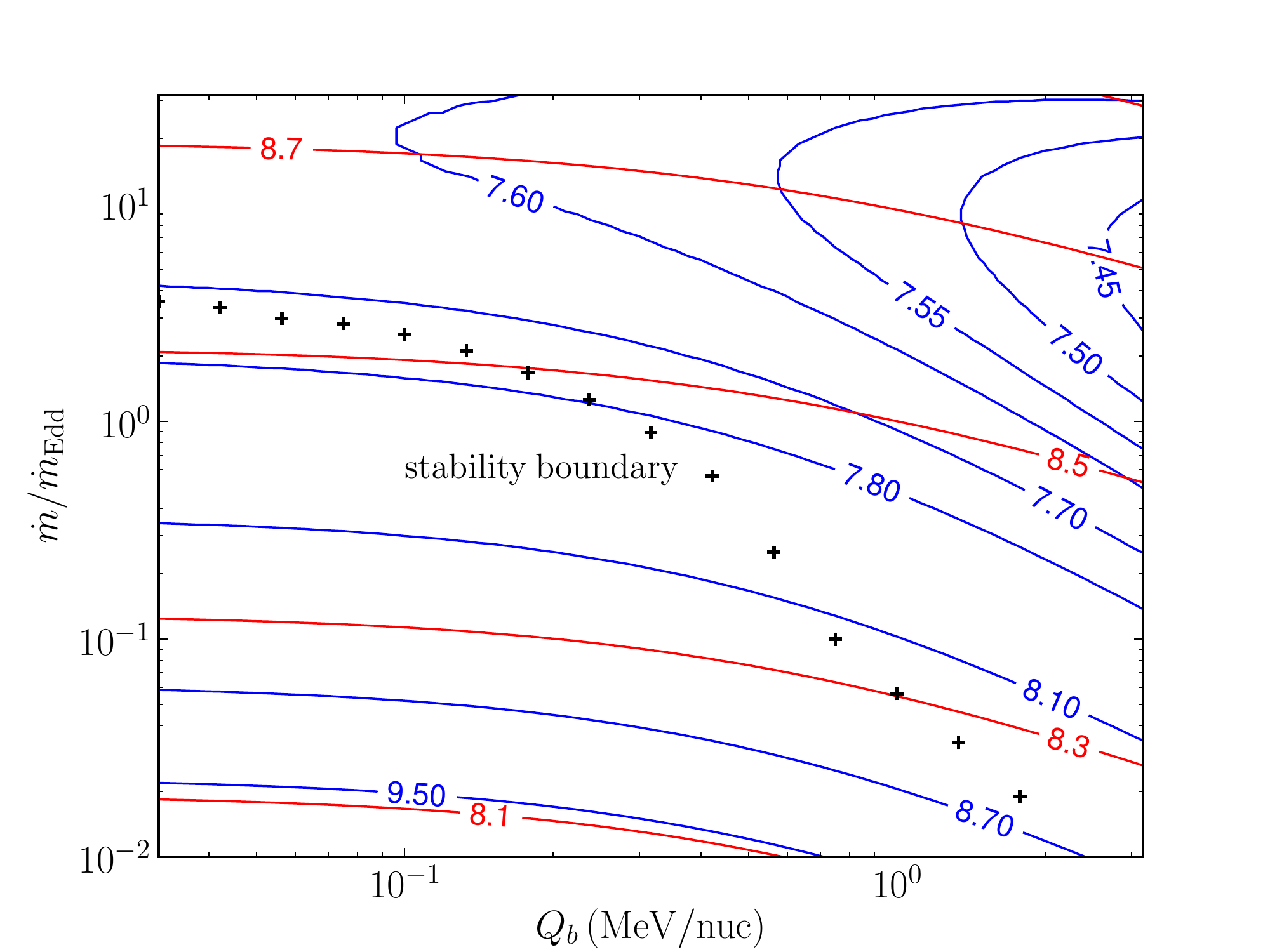}
\caption{Contour lines for constant values of burning column depth ($y_{\rm burn}$; blue lines), that is, the depth at which the triple-$\alpha$ burning rate peaks, and temperature at the burning depth ($T_{\rm burn}$; red lines). The numbered labels on the contours show the base-10 logarithm of the respective quantities. Note that the $\log_{10}T_{\rm burn}=8.7$ contour line appears to very nearly pass through the stationary points of the $y_{\rm burn}$ contour curves, that is, where $dy_{\rm burn}/d\dot{m}=0$. This is addressed in \S\ref{sec:dydmdot_1zone}.
\label{fig:Ty_contours}}
\end{figure}

For our steady-state models, we set the lower boundary at $y=10^{11}\ {\rm g\ cm^{-2}}$. This is deep enough that helium burning is complete at the base. As Figure \ref{fig:Ty_contours} shows, the helium burning depth is typically $10^8\,$g\,cm$^{-2}$, but can reach $10^{10}$\,g\,cm$^{-2}$ at $\dot m\sim1\%\,\dot{m}_{\rm Edd}$ and $Q_b\lesssim 0.1$\,MeV\,nuc$^{-1}$. Our inner boundary also lies above the depth where carbon is likely to burn \citep{Brown1998}. Furthermore, the temperature profile is expected to turn over at some point, with heat being transported into the crust and core. We stop our integrations at a depth shallower than both the temperature turn over point, and the carbon ignition depth. The value of $Q_b$ should be interpreted as the outwards flux evaluated at the lower boundary depth, $y=10^{11}$\,g\,cm$^{-2}$.

The compressional heating gives some sensitivity to the choice of the location of the lower boundary. Beneath the helium burning depth, the layer is close to isothermal, $\nabla$ is much smaller than $\nabla_{\rm ad}$, and $c_PT\nabla_{\rm ad}$ is roughly constant allowing an estimate of the contribution to the flux from compressional heating,
\begin{align}\label{eq:Qcomp}
{dQ_\mathrm{comp}\over d\log_{10}\, y}\,&=\, 0.013\,\frac{\rm MeV}{\rm nuc}\,T_{9}\notag\\
&\times\left(\frac{c_p}{3.0\times10^7\,{\rm erg\ g^{-1}}\,{\rm K}^{-1}}\right)\left(\frac{\nabla_{\rm ad}}{0.4}\right),
\end{align}
where we take a typical value of $c_P$ from our numerical models.
Every additional decade in column depth included below the helium burning depth contributes an extra $0.013$ MeV per nucleon. This means that models with small $Q_b\ll 0.1$ MeV per nucleon actually have a flux heating the helium burning layer that is substantially larger than $Q_b$. In other words, compressional heating in the ocean sets an effective lower limit on the base heating of the helium burning layer. The contributions to the total compressional heat flux are roughly evenly divided between depths below and above the helium burning depth. From equation (\ref{eq:Qcomp}), we estimate the contribution from below to be $\sim0.04$\,MeV/nuc for a typical burning depth and our choice of lower boundary, which gives a total $Q_{\rm comp}=0.08$\,MeV/nuc.

Figure \ref{fig:Ty_contours} shows contours of the burning depth and temperature. We define the burning depth $y_{\rm burn}$ as the location where the 3$\alpha$ burning rate is maximal. The burning temperature $T_{\rm burn}$ is defined as the temperature at the depth $y_{\rm burn}$. The stability boundary as calculated in \S\ref{sec:mesa} is shown. Clearly, $dy_{\rm burn}/d\dot m<0$ along the stability boundary. Interestingly, the locus of points where $dy_{\rm burn}/d\dot m=0$ (where the blue contours turn over) follows closely the temperature contour where $\log_{10}T=8.7$ or $T_8=5$, as the arguments from the one-zone model indicated. We conclude that the correspondance between $dy_{\rm burn}/d\dot m=0$ and marginal stability does not carry over into multizone models for any value of $Q_b$.


\section{Linear stability analysis}\label{sec:linstab}

In this section, we carry out a linear stability analysis of the steady-state models described in \S 3.2. A similar technique was used by \cite{NarayanHeyl2003}, although applied to artificially truncated steady state models in an attempt to calculate ignition conditions in the unstable regime. Here, we are interested in locating the stability boundary and so perturb full steady-state models that burn to completion. This technique should reproduce the stability boundary, since we will identify those values of $\dot m$ and $Q_b$ where the steady-state model is unstable.

We first derive the perturbation equations and boundary conditions in \S \ref{sec:perturb}, and present the results in \S 4.2.

\begin{figure}
\includegraphics[width=0.5\textwidth]{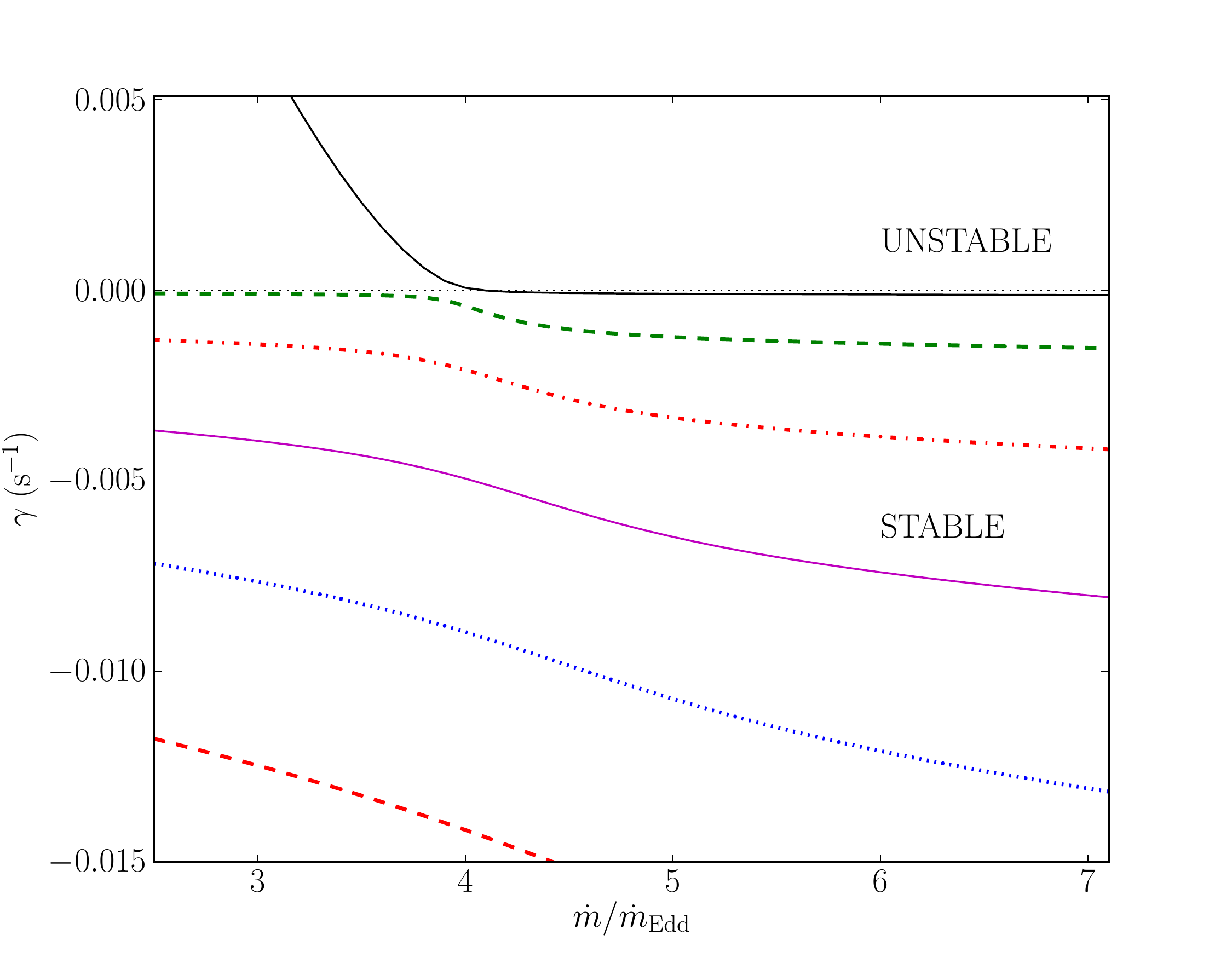}
\caption{The change in eigenvalue $\gamma$ with accretion rate for the first few eigenmodes using a $Q_{base}=0.2$ MeV. The first eigenmode (black line) transitions from unstable ($\gamma>0$) to stable as the accretion rate increases past $\sim4\dot{m}_{\rm Edd}$. The other eigenmodes are stable across all accretion rates. A horizontal dotted black line is used to highlight the location of the transition, $\gamma=0$.
\label{fig:gamma_curves}}
\end{figure}

\subsection{Perturbation equations}\label{sec:perturb}

For the perturbation analysis, we use pressure or equivalently column depth as the independent coordinate (pressure and column depth are related by $P=gy$ in a thin layer, where $g$ is the constant gravity). At each pressure $P$, we set $T\rightarrow T+\delta T$ and $F\rightarrow F+\delta F$, where the perturbations have a time-dependence $e^{\gamma t}$. With the choice of pressure coordinates, we are adopting Lagrangian perturbations. In the Appendix, we derive the perturbation equations from an Eulerian approach, in which vertical displacements are followed explicitly. We assume that on the timescale of the thermal perturbation, the composition does not change $\delta Y=0$, since only a small amount of helium need burn for a large change in temperature (see eqs.~[\ref{eq:dTdt}] and [\ref{eq:dydt}]).

Putting the time-dependent term $c_P\partial T/\partial t$ back into equation (\ref{eq:dFdy}) and perturbing, we find
\begin{equation}\label{eq:ddelFdy}
{d\over dy}\delta F=\left(c_P\gamma-{\epsilon_{3\alpha}\epsilon_T\over T}\right)\delta T
\end{equation}
where $\epsilon_T=\partial\ln\epsilon_{3\alpha}/\partial \ln T|_P$ and we have neglected the compressional heating term. The radiative diffusion equation (\ref{eq:dTdy}) gives 
\begin{equation}\label{eq:ddelTdy}
{d\over dy}\delta T = {dT\over dy}\left({\delta F\over F}+\left[{\kappa_T-3\over T}\right]\delta T\right)
\end{equation}
where $\kappa_T=\partial\ln\kappa/\partial \ln T|_P$. For a given steady-state model ($T(y), Y(y), F(y)$) obtained by integrating equations (\ref{eq:dTdy})--(\ref{eq:dYdy}), the perturbation equations (\ref{eq:ddelFdy}) and (\ref{eq:ddelTdy}) form an eigenvalue problem for $\gamma$, i.e.~the perturbation equations and their boundary conditions will be satisfied only for particular choices of the growth (or decay) rate $\gamma$.

At the top of the layer, the boundary condition comes from perturbing a radiative zero solution ($F = acT^4/3\kappa y$) for the outer layers, giving
\begin{equation}\label{eq:radzero_BC}
{\delta F\over F}=\left(4-\kappa_T\right){\delta T\over T},
\end{equation}
where the choice of $\delta T/T$ at the top is arbitrary and sets the overall normalization. At the base of the layer, the usual approach would be to set $\delta T=0$, which is appropriate when the thermal timescale at the base is much longer than the growth rate of the mode, $\gamma^{-1}$. At marginal stability, however, the growth timescale becomes very long and exceeds the thermal timescale at the base, in which case it is not clear how to set the lower boundary condition. In fact, we find that the results do not depend sensitively on the choice of lower boundary condition. For the results shown, we fix the flux at the base $\delta F=0$. 
We find that changing the boundary condition from $\delta F=0$ to $\delta T=0$ at the base lowers $\dot m_{\rm crit}$ by $<10$\% for $Q_b<0.5$ MeV per nucleon. The differences in $\dot{m}_{crit}$ become larger, roughly a factor of 2, for $Q_b>1$ MeV per nucleon. 

\begin{figure}
\includegraphics[width=0.5\textwidth]{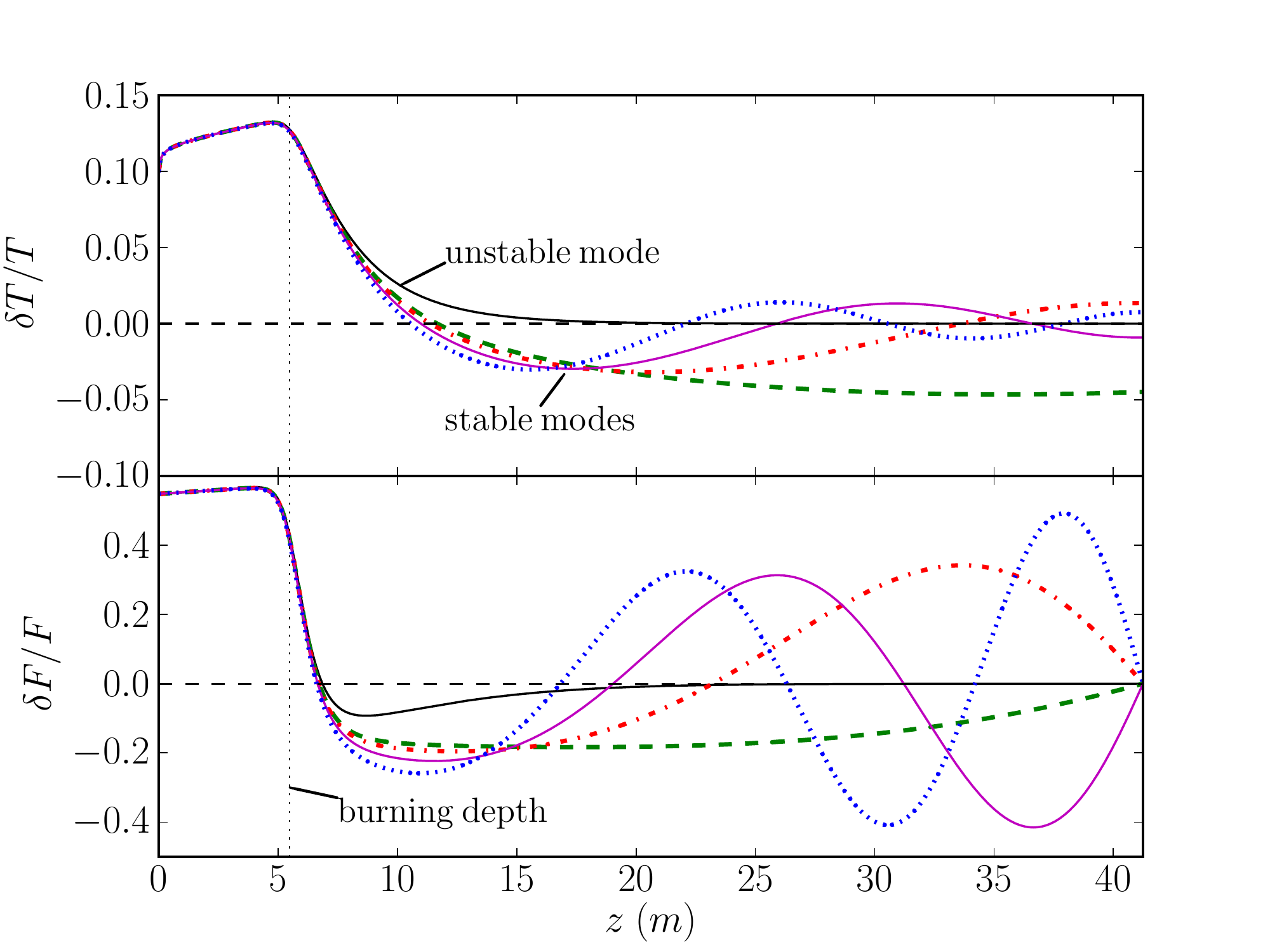}
\caption{Example of the first few perturbed temperature (top panel) and flux (bottom panel) eigenmodes for $\dot{m}\,=\,3\dot{m}_{\rm Edd}$ and $Q_{base}=0.2$ MeV\,nuc$^{-1}$. The eigenmodes displayed have a correspondence with those shown in figure \ref{fig:gamma_curves} at $\dot{m}\,=\,3\dot{m}_{\rm Edd}$, sharing the same line styles and colours. The burning depth, that is, the location at which the triple-alpha burning rate is a maximum in the steady-state models, is represented by the vertical dotted line at roughly $z=5$ m. A horizontal dashed black line is used to highlight the location of $\delta T/T=0$ in the top panel, and $\delta F/F=0$ in the bottom panel. The normalization is chosen so that $\delta T/T = 0.1$ at the top of the layer.
\label{fig:eigenmodes}}
\end{figure}

\subsection{Results}\label{sec:results}

At any given accretion rate and base flux, there are many stable ($\gamma<0$) eigenmode solutions and at most one unstable mode. The unstable mode, if present, transitions to stability at a specific accretion rate --- this defines the stability boundary. As an example, Figure \ref{fig:gamma_curves} shows the values of $\gamma$ as a function of $\dot m$ for the first six eigenmodes, for a base flux $Q_b=0.2$ MeV per nucleon. The lowest order mode has $\gamma>0$ (unstable) for $\dot m\gtrsim 4\ \dot m_{\rm Edd}$ and $\gamma<0$ (stable) for $\dot m\lesssim 4\ \dot m_{\rm Edd}$. Figure \ref{fig:eigenmodes} shows the eigenmodes at $\dot m=3\ \dot m_{\rm Edd}$, just below the stability boundary in the unstable region, again for $Q_b=0.2$ MeV per nucleon. The unstable mode has a single peak in $\delta T$ at the triple-$\alpha$ burning depth, since this is the location at which the thermal runaway occurs during the onset of a burst. The stable (cooling) modes show oscillations, with an increasing number of nodes associated with decreasing (larger negative) values of $\gamma$. The cooling eigenmodes with the most negative $\gamma$ decay most quickly, due to the $e^{\gamma t}$ time dependence of the perturbations.

In \citet{NarayanHeyl2003}, a slightly different instability criterion was used, namely that the unstable mode growth timescale $1/\gamma$ be shorter than three times the accretion timescale $y_{\rm ign}/\dot m$. The values of $\dot m_{\rm crit}$ we found with this criterion were not substantially different from those found using $\gamma>0$. For example, in the case of $Q_b=0.2$\,MeV per nucleon, the \citet{NarayanHeyl2003} instability criterion (here, roughly $\gamma>10^{-3}$\,s$^{-1}$) leads to a $\lesssim10\%$ decrease in the value of $\dot m_{\rm crit}$. As is shown in Figure~\ref{fig:gamma_curves}, this modest decrease can be understood from the sudden steep rise in $\gamma$ with decreasing $\dot m$ for the dominant mode (solid black curve), below $\dot m/\dot m_{\rm Edd}\simeq4$.

\begin{figure}
\includegraphics[width=0.5\textwidth]{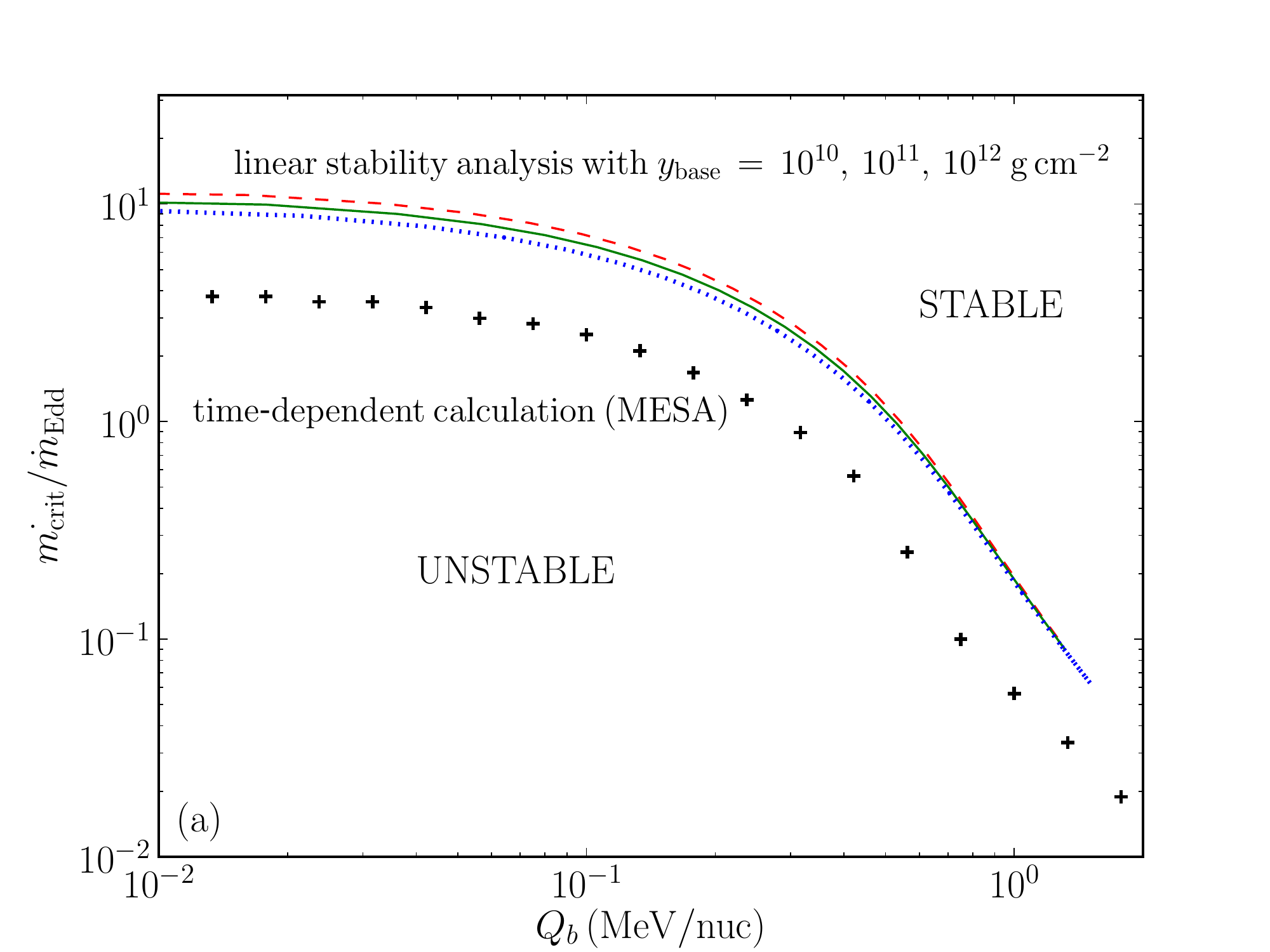}
\includegraphics[width=0.5\textwidth]{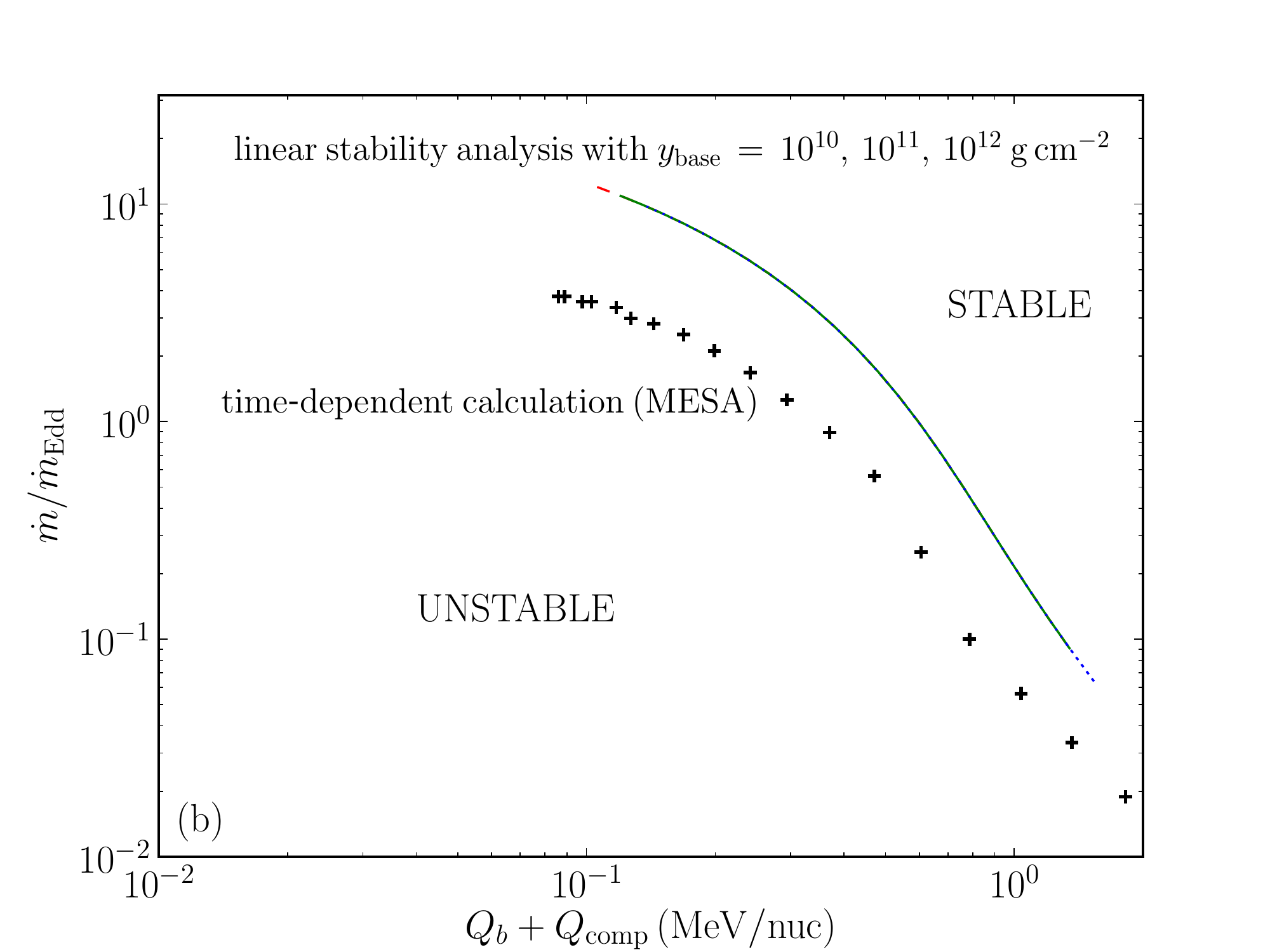}
\caption{(a) The stability boundary for helium burning as found using linear stability analysis (\S\ref{sec:linstab}) is shown as a green solid curve. In addition, the stability curves for the same calculation are shown, using different lower boundary depths, namely $y_{\rm base}=10^{10}$\,g\,cm$^{-2}$ (red dashed), $y_{\rm base}=10^{12}$\,g\,cm$^{-2}$ (blue dotted). The MESA stability boundary is represented by black crosses. (b) The same stability boundaries shown in (a), this time plotted against $Q_{b}+Q_{\rm comp}$, the sum of $Q_b$ and the total contribution to compressional heating across the layer. The linear stability analysis curves with different base depths now overlap each other.
\label{fig:Qbmdot}}
\end{figure}

Rather than searching for $\gamma=0$ by varying $\dot m$, we locate $\dot m_{\rm crit}$ at each $Q_b$ by setting $\gamma=0$ and then treating $\dot m$ as the eigenvalue. The resulting stability boundary is shown in the top panel of Figure \ref{fig:Qbmdot}, as the solid green curve. We have also included stability curves for the same calculation but with different lower boundaries, $y_{\rm base}=10^{10}$ and $10^{12}\ {\rm g\ cm^{-2}}$. This illustrates the effect of compressional heating: a deeper layer has additional compressional heating, increasing the flux heating the helium layer and stabilizing the burning, moving $\dot m_{\rm crit}$ to lower values. As can be seen, the effect is not large, with a $\approx 20$\% change in $\dot m_{\rm crit}$ over the factor of 100 change in $y_{\rm base}$.

To correct for the effect of compressional heating on the stability boundary, in the lower panel of Figure \ref{fig:Qbmdot} we show $\dot m_{\rm crit}$ against the sum of $Q_b$ and $Q_{\rm comp}$, which gives the total flux heating the helium burning layer, plus a contribution to $Q_{\rm comp}$ coming from depths shallower than helium burning. In other words, $Q_b+Q_{\rm comp}$ represents the total non-nuclear heating flux which emerges at the top of the accreted layer. The linear stability curves now lie on top of one another for all choices of $y_{\rm base}$. This plot also emphasizes that compressional heating in the ocean sets a minimum value for the effective base flux of $\sim0.1$\,MeV/nuc, similar to the estimate of the total compressional heating that we found in \S\ref{sec:steady}.

Recall that in arriving at equation (\ref{eq:ddelFdy}), we did not include perturbations of the compressional heating terms from equation (\ref{eq:dFdy}). We checked the effect of including these terms on the stability boundary, and found only a small $6-7\%$ increase in the value of the critical accretion rate. In addition, we also evaluated the effect of changing the surface gravity. A surface gravity of $g_{14}=1.0$\,cm\,s$^{-2}$ yielded an $\sim30\%$ decrease in the value of the critical accretion rate, while $g_{14}=3.0$\,cm\,s$^{-2}$ yielded a $\sim25\%$ increase. These results agree very well with the $\dot m_{\rm crit}\propto g_{14}^{1/2}$ scaling found by \citet{Bildsten1998}.

\begin{figure}
\includegraphics[width=0.5\textwidth]{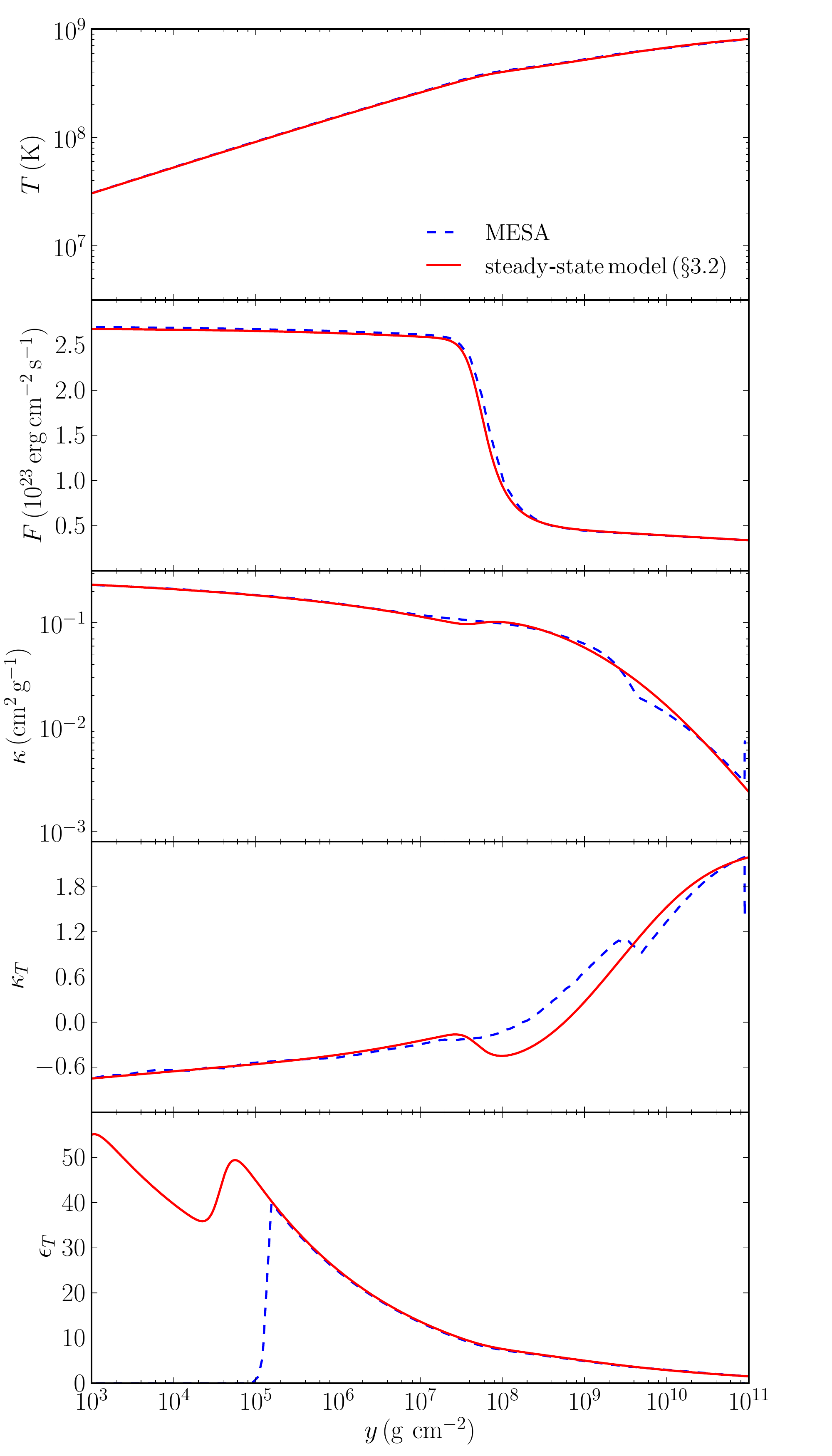}
\caption{A comparison of steady burning model calculations using MESA (blue) and the model presented in \S\ref{sec:steady}, with $\dot{m}=4\dot{m}_{\rm Edd}$ and $Q_{b}=0.1$\,MeV\,nuc$^{-1}$. At this accretion rate and base flux, the MESA simulation indicates that this model is stable, while our linear stability analysis (\S\ref{sec:perturb}) indicates that the model is unstable. Profiles for the temperature, flux, opacity, and opacity and burning rate derivatives $\kappa_T$, $\epsilon_T$ (both taken at constant pressure) are shown.\label{fig:profiles}}
\end{figure}

Figure \ref{fig:Qbmdot} shows that the $\dot m_{\rm crit}$ calculated with linear stability analysis is a factor of $\approx 3$ greater than the $\dot m_{\rm crit}$ determined from the time-dependent MESA simulations. The reason for this discrepancy is not clear. We have compared our steady-state models with MESA for values of $\dot m$ and $Q_b$ at which MESA achieves a steady solution, while our linear stability analysis predicts instability, and find excellent agreement (see Fig.~\ref{fig:profiles}). There are small differences in the opacity profile and $\kappa_T$ (lower panel of Fig.~\ref{fig:profiles}), but these differences make only a small change in the growth rate. For example, we calculated the linear growth rate for a model with $\dot m = 4\dot m_{\rm Edd}$ and $Q_b=0.1\ {\rm MeV}$ per nucleon using the $\kappa_T$ profile from MESA, and compared it to the growth rate found using the $\kappa_T$ profile from our steady-state models, but found only a small difference, $\gamma=0.016\ {\rm s^{-1}}$ compared to $\gamma=0.018\ {\rm s^{-1}}$. Therefore the difference in $\kappa_T$ profile is not the reason that the model is stable in MESA but unstable according to the linear stability analysis. Lastly, while the profiles for $\epsilon_T$ diverge dramatically at depths $y\lesssim10^5$\,g\,cm$^{-2}$, this parameter is irrelevant at these depths since the helium burning rate is negligible. Another possible reason for the difference could be that we have not allowed changes in composition in our linear stability analysis, setting $\delta Y=0$. However, we do not expect these extra terms to significantly change the results, since the thermal timescale is shorter than the timescale to change composition by a factor $\approx Q_{3\alpha}/C_PT\approx 10$ for $T_8\approx 3$.

\section{Summary and Discussion}

The main result of the paper is a new calculation of the critical accretion rate $\dot m_{\rm crit}$ at which helium burning stabilizes on accreting neutron stars. We used the MESA stellar evolution code to calculate $\dot m_{\rm crit}$ as a function of the base flux heating the helium layer, written in terms of the energy per nucleon $Q_b$ ($F=\dot m Q_b$). Equation (\ref{eq:mdotcrit}) gives an analytic expression for $\dot m_{\rm crit}(Q_b)$ in units of the local Eddington rate $\dot m_{\rm Edd}=8.8\times 10^4\ {\rm g\ cm^{-2}\ s^{-1}}$, which should be useful for applications.

In agreement with \cite{Keek2009}, we find that the critical accretion rate at low fluxes, $\dot m_{\rm crit}\approx 4\ \dot m_{\rm Edd}$ is substantially smaller than the rate $\dot m_{\rm crit}\approx 20\ \dot m_{\rm Edd}$ predicted by one-zone models \citep{Bildsten1995,Bildsten1998}. The difference arises because the one-zone instability criterion (eq.~[\ref{eq:instability_criterion}]) overestimates the burning temperature at marginal stability, which is close to $3.5\times 10^8\ {\rm K}$ in multizone models but predicted to be $5\times 10^8\ {\rm K}$ in a one-zone model. 

We also investigated whether the critical accretion rate can be determined by examining steady-state models only, without running a time-dependent simulation. \cite{Paczynski1983a} showed that a one-zone model with $Q_b=0$ has a turning point in the burning depth $dy/d\dot m=0$ at marginal stability. We find that this result does not hold in one-zone models when a base flux is included, and does not hold for any value of $Q_b$ in multizone models. This is contrary to the findings of \cite{NarayanHeyl2003}, who studied multizone models with fixed temperatures as a lower boundary.

We then carried out a linear stability analysis of steady-state burning models to determine the stability boundary. Linear stability analysis has been applied to nuclear burning on neutron stars before (e.g.~\citealt{NarayanHeyl2003}), but not compared directly to time-dependent simulations. Although the shape of the $\dot m_{\rm crit}(Q_b)$ curve is reproduced quite well (Fig.~7), the linear stability analysis overestimates $\dot m_{\rm crit}$ by a factor of about 3. We were not able to identify the reason for the discrepancy; for now we must take the results of linear stability analysis as approximate. \cite{NarayanHeyl2003} assumed a solar composition, and so cannot be compared with our results.

\citet{Heger2007} discussed a further prediction of theoretical models, that close to marginal stability, the eigenvalue of thermal perturbations becomes complex \citep{Paczynski1983a}, leading to an oscillatory mode of burning which has been identified with mHz frequency quasi-periodic oscillations (mHz QPOs) observed from 3 X-ray binaries \citep{Revnivtsev2001,Altamirano2008,Linares2012}. By considering only thermal perturbations in this paper, we have confined our attention to the real part of the eigenvalue, neglecting the compositional perturbations that are important in marginally-stable burning. This is a straightforward extension of the method presented here, and remains to be addressed in a future paper.

It would be interesting to apply the linear stability analysis to steady-state models of solar composition, which include hydrogen burning by the rp-process. A large grid of models were recently published as a function of $Q_b$ and helium fraction $Y$ \citep{Stevens2014}. \cite{Heger2007QPOs} found the stability boundary $\dot m_{\rm crit}=0.924\ \dot m_{\rm Edd}$ for $Q_b=0.15\ {\rm MeV}$ per nucleon in simulations with the KEPLER code. \cite{Keek2014} extend these calculations to investigate the sensitivity of $\dot m_{\rm crit}$ to nuclear reaction uncertainties. For their standard set of rates, they have $\dot m_{\rm crit}\approx 1.1\ \dot m_{\rm Edd}$. \cite{Bildsten1998} estimated $\dot m_{\rm crit}=0.74\ \dot m_{\rm Edd}$ for solar composition using the one-zone ignition criterion. In that case, the one-zone estimate appears to give a much more accurate estimate than for pure helium. 

The transition to stable burning is believed to explain the observed quenching of Type I X-ray bursts following a superburst \citep{Kuulkers2002,Bildsten2001,Cumming2004,Keek2012}. \cite{Cumming2004} assumed that the critical flux that would quench burning is $Q_b\approx 0.7\ {\rm MeV}$ per nucleon, independent of accretion rate. In fact, as we showed in this paper, we expect the $Q_b$ required to stabilize burning to depend strongly on $\dot m$. Superburst sources are not pure helium accretors in general, but we can compare our results with \cite{Keek2012}, who ran time-dependent simulations of superbursts and studied quenching for the pure helium case. They found that burning became unstable as the luminosity dropped through $L\approx 4\times 10^{35}\ {\rm erg\ s^{-1}}$ for accretion at 0.3 $\dot m_{\rm Edd}$. Subtracting the nuclear burning flux, this is in good agreement with Figure \ref{fig:Qbmdot} which predicts a critical flux of $Q_b\approx 0.5\ {\rm MeV}$ per nucleon for this accretion rate. The fact that this is close to the value assumed by \cite{Cumming2004} suggests that their results may not be strongly affected by their assumption that $Q_b$ is independent of $\dot m$.

Our results can be immediately applied to 4U~1820-30, an ultracompact binary that most likely accretes pure helium. It displays regular Type I X-ray bursts in its low state, which disappear when the accretion rate increases and the source enters the soft state \citep{Clark1977,Cornelisse2003}. \citet{Cumming2003} found that at the local rate of $\dot m_X=1.2\times 10^4\ {\rm g\ cm^{-2}\ s^{-1}}=0.14\ \dot m_{\rm Edd}$ (as inferred from the X-ray luminosity of the source when bursts are seen), a flux from below of $Q_b=0.4\ {\rm MeV}$ per nucleon was necessary to explain the short $\approx 3$ hours burst recurrence times.
For this value of $Q_b$, we find that burning will stabilize above $\dot m=0.35\ \dot m_{\rm Edd}$ (using eq.~[\ref{eq:mdotcrit}]). This can be accommodated in the range of accretion rates observed in the 6 month cycle of 4U~1820-30, which is about a factor of 3. Therefore, it may be possible to make a consistent model of the burst recurrence time and the quenching of bursts at higher accretion rates by including a base flux of the appropriate size that is always present. An alternative is that the flux switches on at a critical rate, quenching the burning, but this would have difficulty explaining the short recurrence times when bursts are seen. Time-dependent simulations, e.g.~with the MESA code, are required to test whether a self-consistent model of the bursting behavior of 1820-30 can be made.

One issue for explaining the transition to stable burning is the timescale on which bursts appear or disappear as the accretion rate changes. \cite{Zand2012} noted that the burst behavior in 4U~1820-30 changes within a day or two of entering or leaving the low state. They suggest that this implies that the shallow heat source must lie at a depth where the thermal time is $\lesssim 1$ day, corresponding to a density of $\rho\approx 10^9\ {\rm g\ cm^{-2}}$, so that it can adjust to the changing accretion rate. Otherwise, for example, when the accretion rate dropped into the low state, the luminosity from the crust would remain as it was in the high state, not having time to thermally adjust, and X-ray bursts would remain quenched. Instead, we want the luminosity to adjust to a new value of $Q_b\dot m$ so that bursting activity can resume.  

The fact that the stability boundaries for pure helium and solar composition are closer than previously thought (based on one-zone models, in which they are more than an order of magnitude different) may help to explain why burning stabilizes in 4U~1820-30 at a similar accretion rate to other low mass X-ray binary neutron stars that accrete hydrogen rich material. 

\vspace{10mm}
We thank L. Keek and G. Ushomirsky for useful discussions. We are grateful for support from the National Sciences and Engineering Research Council (NSERC) of Canada. AC is an Associate Member of the CIFAR Cosmology and Gravity program. MZ \& AC are members of an International Team in Space Science on thermonuclear bursts sponsored by the International Space Science Institute in Bern, Switzerland. 

\bibliographystyle{mn2e}
\bibliography{refs}

\begin{appendix}

\section{Eulerian perturbations}\label{appendix:eulerian}

In \S 4.1, we derived the perturbation equations using pressure coordinates, a Lagrangian approach. Here we instead use an Eulerian approach, where perturbations are taken at fixed spatial position, and show that the perturbation equations reduce to those derived in \S 4.1 when written in terms of Lagrangian quantities. We follow the convention of \citet{Cox1980} by denoting Eulerian perturbations using the prime symbol. For example, $T'$ represents the Eulerian temperature perturbation. The Lagrangian temperature perturbation is then $\delta T = T^\prime + \xi_z \partial T/\partial z$, where $\xi_z$ is the vertical displacement. The displacement obeys the continuity equation
\begin{equation}\label{eq:dxidy}
{d\over dy}\xi_z \,= {\delta \rho\over \rho^2}  = \,-{\chi_T\over\rho\chi_\rho}\frac{\delta T}{T},
\end{equation}
where we have set $\delta P=0$.

Perturbing equation (\ref{eq:dTdy}) using Eulerian perturbations gives
\begin{equation}\label{eq:Aa}
-\frac{1}{\rho}\frac{\partial T'}{\partial z}\ =\ -\frac{1}{\rho}\frac{\partial T}{\partial z}\left[\frac{F'}{F} -3\frac{T'}{T} + \frac{\kappa'}{\kappa} + \frac{\rho'}{\rho} \right].
\end{equation}
Now to rewrite this in terms of Langrangian perturbations. The gradient of the Lagrangian temperature perturbation is 
\begin{equation}
\frac{\partial\delta T}{\partial z}\ =\ \frac{\partial T'}{\partial z} + \frac{\partial}{\partial z}\left(\xi_z\frac{\partial T}{\partial z} \right) =\ \frac{\partial T'}{\partial z} -\frac{\partial T}{\partial z}\frac{\delta \rho}{\rho} + \xi_z\frac{\partial^2T}{\partial z^2},
\end{equation}
where we used the continuity equation $d\xi_z/dz\,=\, -\delta \rho/\rho$ to substitute for $\xi_z$. Combining this with equation (\ref{eq:Aa}) gives
\begin{align}
\frac{\partial\delta T}{\partial z}\ &=\ \frac{\partial T}{\partial z}  \left[\frac{\delta F}{F} -3\frac{\delta T}{T} + \frac{\delta \kappa}{\kappa} \right] \notag\\
& -\xi_z\frac{\partial T}{\partial z} \frac{\partial}{\partial z}\left[ \ln\left( \frac{F \kappa \rho}{T^3 \frac{\partial T}{\partial z}}\right) \right].
\end{align}
The last term in equation (A3) vanishes since the expression inside the logarithm is a constant, giving
\begin{equation}
\frac{\partial\delta T}{\partial y}\ =\ {dT\over dy}\left[{\delta F\over F}+\left({\kappa_T-3\over T}\right)\delta T\right].
\end{equation}
We have recovered equation (\ref{eq:ddelTdy}) from \S 4.1.

Next, the Eulerian-perturbed entropy equation is
\begin{equation}\label{eq:eul2}
\gamma c_P \delta T\ =\ \epsilon'  +\frac{1}{\rho}\frac{\partial F}{\partial z}\frac{\rho'}{\rho} -\frac{1}{\rho}\frac{\partial F'}{\partial z}.
\end{equation}
As above, we express the Eulerian perturbations as Lagrangian perturbations:
\begin{align}
\gamma c_P \delta T \ &=\ \delta\epsilon  +\frac{1}{\rho}\frac{\partial F}{\partial z}\frac{\delta \rho}{\rho}  -\frac{1}{\rho}\frac{\partial \delta F}{\partial z}  + \frac{1}{\rho}\frac{\partial F}{\partial z}\frac{\partial\xi_z}{\partial z} \notag\\
 & -\xi_z\left[\frac{\partial \epsilon}{\partial z} + \frac{1}{\rho}\frac{\partial F}{\partial z}\frac{d\ln\rho}{dz} -\frac{1}{\rho}\frac{\partial}{\partial z}\frac{\partial F}{\partial z}  \right].
\label{eq:A13}
\end{align}
Using the expression $\delta \rho/\rho\,=\,-d\xi_z/dz$, and
\begin{equation}
\frac{\partial}{\partial z}\left(\frac{1}{\rho}\frac{\partial F}{\partial z} \right)\ =\ \frac{1}{\rho}\frac{\partial}{\partial z}\frac{\partial F}{\partial z} - \frac{1}{\rho}\frac{\partial \ln\rho}{\partial z}\frac{\partial F}{\partial z},
\end{equation}
equation (\ref{eq:A13}) simplifies to
\begin{equation}
\gamma c_P \delta T \ =\ \delta\epsilon  -\frac{1}{\rho}\frac{\partial \delta F}{\partial z} -\xi_z\frac{\partial}{\partial z}\left[ \epsilon -\frac{1}{\rho}\frac{\partial F}{\partial z}  \right].
\end{equation}
The two terms inside the bracket cancel out in steady state, and we are left with
\begin{equation}
\frac{\partial \delta F}{\partial y}\ =\ \delta T\left(\gamma c_P  -\frac{\epsilon\epsilon_T}{T}\right),
\end{equation}
which is equation (\ref{eq:ddelFdy}) from \S 4.1.

The set of Eulerian perturbed equations (\ref{eq:dxidy}), (\ref{eq:Aa}), and (\ref{eq:eul2}) are physically equivalent to the Lagrangian perturbation equations. If we use the same boundary conditions, as outlined in \S 4.1, with the additional condition on the vertical displacement, $\xi_z = 0$ at the base, we get the same solutions. However, integration of the Eulerian equations is more complex computationally, because of the additional boundary condition.

\end{appendix}

\label{lastpage}

\end{document}